\documentclass[12pt]{article}
\newcommand{\bfP}{{\bf P}}
\newcommand{\RRR}{$I\hspace{-0.25em}R^3$}
\newcommand{\ket}[1]{\vert#1\rangle}

\newcommand{\braket}[2]{\langle#1\vert#2\rangle}
\newcommand{\ketbra}[2]{\vert#1\rangle\langle#2\vert}

\newcommand{\bd}{\begin{displaymath}}
\newcommand{\ed}{\end{displaymath}}
\newcommand{\be}{\begin{equation}}
\newcommand{\ee}{\end{equation}}
\newcommand{\bi}{\begin{itemize}}
\newcommand{\ei}{\end{itemize}}
\newcommand{\bq}{\begin{quote}}
\newcommand{\eq}{\end{quote}}

\begin{document}
\setlength{\baselineskip}{16.5pt}
\title{Do quantum states evolve?\\
Apropos of Marchildon's remarks}
\author{Ulrich Mohrhoff\\
Sri Aurobindo International Centre of Education\\
Pondicherry 605002 India\\
\normalsize\tt ujm@auromail.net}
\date{}
\maketitle
\begin{abstract}
\noindent Marchildon's (favorable) assessment (quant-ph/0303170, to appear in Found. 
Phys.) of the Pondicherry interpretation of quantum mechanics raises several issues, which 
are addressed. Proceeding from the assumption that quantum mechanics is fundamentally a 
probability algorithm, this interpretation determines the nature of a world that is irreducibly 
described by this probability algorithm. Such a world features an objective fuzziness, which 
implies that its spatiotemporal differentiation does not ``go all the way down''. This result is 
inconsistent with the existence of an evolving instantaneous state, quantum or otherwise.
\setlength{\baselineskip}{14pt}
\end{abstract}

\vspace{\baselineskip}\section{\large INTRODUCTION}
Louis Marchildon~\cite{Marchildon} has undertaken a difficult task: to make sense of my 
published attempts to make sense of quantum mechanics. (To minimize the use of ``I'' and 
``my'' I will refer to them collectively as the ``Pondicherry interpretation of quantum 
mechanics'', PIQM~\cite{piqm}.) On the whole he has succeeded admirably, notwithstanding 
the difficulty of the task, which is compounded by the fact that our neurobiology militates not 
only against attempts to make sense of the quantum world~\cite{bccp,clicks,pseudo} but 
also against attempts to communicate the way that, according to the PIQM, the world is. 
Here I respond to some of the issues raised by Marchildon's assessment of the PIQM.

Marchildon mentions two approaches to the so-called ``pointer problem''---decoherence 
theories~\cite{JoosZeh,Zurek} and modal interpretations~\cite{VF,Dieks3,Vermaas}---and 
states that the PIQM ``does not seem to bring any additional insight to the issue''. I think it 
does. By his own admission, his ``critical examination of a number of assertions'' made by 
the PIQM does ``not cover all aspects of the interpretation, which partakes of a wide-ranging 
system that reaches well into metaphysical ontology''. It appears to me that Marchildon has 
missed an aspect of the PIQM that is both metaphysical and relevant to the pointer problem. 
This aspect, explained in Sec.~2, is the limited (or finite) spatiotemporal differentiation of the 
physical world, which follows from the relative and contingent reality of the spatial 
distinctions that we make, and which is inconsistent with the existence of an evolving 
instantaneous state. Its bearing on the pointer problem is discussed in Sec.~3.

According to the PIQM, value-indicating events (``measurements'') are uncaused. 
Marchildon considers this conclusion epistemologically ``risky''. Section~4 assesses the risk. 
Section~5 examines two ``good'' reasons adduced by Marchildon for viewing the state 
vector as specifying an evolving (instantaneous) state, and offers further reasons for not 
doing so. Section~6 addresses a couple of specific questions raised by Marchildon and clears 
up an ambiguity concerning the ``time of measurement''. In Sec.~7 the following question is 
addressed: What changes in our concepts of ``state'' and ``evolution'' are needed so as to 
make them applicable to a world that is fundamentally and irreducibly described by the 
probability distributions of~QM? In the concluding section I show how the PIQM reconciles 
two apparently conflicting statements: According to Marchildon, the PIQM ``takes quantum 
mechanics to be fundamental and complete, and it requires the validity of classical 
mechanics for its formulation''.

\section{\large THE FINITE SPATIOTEMPORAL DIFFERENTIATION\\
OF THE PHYSICAL WORLD}
\label{fstd}
The PIQM assumes that QM is {\it fundamentally\/} exactly what somehow or other it 
obviously is---a probability algorithm. It seeks to determine the nature of a world that is 
fundamentally and irreducibly described by this probability algorithm. This must be confusing 
to many. Have we not been told time and again that if QM is nothing but a probability 
algorithm then it is an epistemic theory concerned with ``states of knowledge'' rather than 
``states of nature'', and if it is an ontological theory then the quantum state represents an 
evolving instantaneous state and not just a probability measure?

There are no sufficient grounds for either conclusion. If our fundamental physical theory is an 
algorithm for assigning probabilities to the possible outcomes of possible measurements on 
the basis of actual outcomes, or (what comes to the same) if it encapsulates correlations 
between value-indicating events (VIEs), then these correlations are descriptive of an 
objective fuzziness.%
\footnote{``Uncertainty'' mistranslates the German original ``Unsch\"arfe'', the literal 
meaning of which is ``fuzziness''.}
As a result of this fuzziness, the spatiotemporal differentiation of the physical world is finite; 
it does not ``go all the way down''. But if the world is not infinitely differentiated timewise, 
there is no such thing as an evolving instantaneous state. Or so I will argue in this section.

Consider the probability distribution $|\psi(x)|^2$ associated with the position of the 
electron relative to the nucleus in a stationary state of atomic hydrogen. Imagine a small 
region~$V$ for which $\int_V|\psi(x)|^2dx$ differs from both 0 and~1. While the atom is in 
this state, the electron is neither inside~$V$ nor outside~$V$. (If the electron were inside, 
the probability of finding it outside would be~0, and vice versa.) But being inside~$V$ and 
being outside~$V$ are the only relations that can possibly hold between the electron 
and~$V$. (An unextended object cannot have a part inside~$V$ and another part 
outside~$V$, nor is the position of an object the kind of thing that can have parts.) If neither 
of the possible relations between the electron and the (imagined) region~$V$ hold, this 
region simply does not exist for the electron. It has no reality as far as the electron is 
concerned.

Conceiving of a region~$V$ is tantamount to making the distinction between ``inside~$V$'' 
and ``outside~$V$''. Hence instead of saying that $V$ does not exist for the electron, we 
may say that the distinction we make between ``inside~$V$'' and ``outside~$V$'' is a 
distinction that the electron does not make. Or we may say that the distinction we make 
between ``the electron is inside~$V$'' and ``the electron is outside~$V$'' is a distinction 
that Nature does not make. It corresponds to nothing in the physical world.

Suppose, again, that the observables $A$, $B$, and $C$ are consecutively measured, the 
result of the first measurement being~$a$. In the simplest case in which the Hamiltonian 
is~0, the probability of finding $c$ after the intermediate measurement of $B$ is
\be
\label{e0}
p\,(c|a,B)=\sum_i\left|\braket c{b_i}\braket{b_i}a\right|^2.
\ee
If the Hamiltonian is not~0, the brackets are transition amplitudes. Formula (\ref{e0}) 
applies whenever information concerning the value of~$B$ is in principle available, however 
hard it may be for us to obtain it. If such information is strictly unavailable, the probability of 
finding $c$ is
\be
p\,(c|a)=\left|\braket ca\right|^2=\left|\sum_i\braket c{b_i}\braket{b_i}a\right|^2.
\ee
Thus if the intermediate measurement is made, we add the probabilities associated with the 
alternatives defined by the results of the intermediate measurement. Otherwise we add the 
amplitudes $\braket c{b_i}\braket{b_i}a$ associated with these alternatives. Why?

Because in one case the distinctions we make between the alternatives have counterparts in 
the physical world, and in the other case they don't. To cite a familiar example, in one case 
the electron~($e$) goes through either the left slit~($L$) or the right slit~($R$). The 
propositions ``$e$~went through~$L$'' and ``$e$~went through~$R$'' possess truth 
values; one of them is true, the other false. In the other case, these propositions lack truth 
values; they are neither true nor false but meaningless. (So, therefore, is the proposition 
``$e$~went through both slits'', since the conjunction of two meaningless proposition is 
another meaningless proposition.) Nor does an electron (or the position of a C$_{60}$ 
molecule, for that matter~\cite{Arndtetal}) have parts that go through different slits. The 
electron goes through $L\&R$, the region defined by the slits considered as a whole, but it 
goes neither through~$L$ nor through~$R$ because the distinction involved is a distinction 
that Nature does not make in this case.

Whenever QM requires us to add amplitudes, the distinctions we make between the 
corresponding alternatives are distinctions that Nature does not make. They correspond to 
nothing in the physical world. Hence the reality of spatial distinctions is relative and 
contingent---``relative'' because the distinction we make between the inside and the outside 
of a region may be real for a given object at a given time, and it may have no reality for a 
different object at the same time or for the same object at a different time; and 
``contingent'' because the existence of a given region~$V$ for a given object~$O$ at a 
given time~$t$ depends on whether the proposition ``$O$~is in~$V$ at the time~$t$'' has 
a truth value.

Does this proposition have a truth value if none is indicated (that is, if none can be inferred 
from an actual event or state of affairs)? Here is a related question: Suppose that $W$ is a 
region disjoint from~$V$, and that $O$'s presence in~$V$ is indicated. Isn't $O$'s absence 
from~$W$ indicated at the same time? Are we not entitled to infer that the proposition 
``$O$~is in~$W$'' has a truth value (namely, ``false'')? Because the reality of spatial 
distinctions is relative and contingent, the answer is negative. Regions of space do not exist 
by themselves. The distinction we make between ``inside~W'' and ``outside~W'' has no 
physical reality {\it per se\/}. If $W$ is not realized (made real) by some means, it does not 
exist. But if it does not exist, the proposition ``$O$~is in~$W$'' cannot have a truth value. 
All we can infer from $O$'s indicated presence in~$V$ is the truth of a {\it counterfactual\/}: 
if~$W$ were the sensitive region of a detector~$D$, $O$~would not be detected by~$D$. 
Probability~1 is {\it not\/} sufficient for ``is'' or ``has''.

It follows that a detector%
\footnote{A perfect detector, to be precise. If $D$ is less than 100~percent efficient, the 
absence of a click does not warrant the falsity of ``$O$~is in~$W$''.}
performs two necessary functions: it indicates the truth value of a proposition of the form 
``$O$~is in~$W$'', and by realizing~$W$ (or the distinction between ``inside~$W$'' and 
``outside~$W$'') it makes the predicates ``inside~$W$'' and ``outside~$W$'' available for 
attribution to~$O$. The apparatus that is presupposed by every quantum-mechanical 
probability assignment is needed not only for the purpose of indicating the possession, by a 
material object, of a particular property (or the possession, by an observable, of a particular 
value) but also for the purpose of realizing a set of properties or values, which thereby 
become available for attribution.%
\footnote{This does not mean that QM~is restricted ``to be exclusively about piddling 
laboratory operations''~\cite{BellAM}. Any event from which either the truth or the falsity of 
a proposition of the form ``system~$\cal S$ has property~$P$'' can be inferred, qualifies as 
a measurement.}

Now let \RRR($O$) be the set of unpossessed and purely imaginary exact positions relative 
to an object~$O$. Since no material object ever has a sharp position, we can conceive of a 
partition of \RRR($O$) into finite regions that are so small that none of them is the sensitive 
region of an actually existing detector. Hence we can conceive of a partition of \RRR($O$) 
into sufficiently small but finite regions~$V_i$ of which the following is true: there is no 
object~$Q$ and no region~$V_i$ such that the proposition ``$Q$~is inside~$V_i$'' has a 
truth value. In other words, there is no object~$Q$ and no region~$V_i$ such that $V_i$ 
exists for~$Q$. But a region of space that does not exist for any material object, does not 
exist---period. The regions~$V_i$ represent spatial distinctions that Nature does not make. 
They correspond to nothing in the physical world. The bottom line: The world is not infinitely 
differentiated spacewise. Its spatial differentiation is finite---it doesn't go all the way down.

While positions are indicated by (macroscopic) detectors, times are indicated by 
(macroscopic) clocks. (``Macroscopic'' is defined in the following section.) Since clocks 
indicate times by the positions of their hands,%
\footnote{Digital clocks indicate times by transitions from one reading to another, without 
hands. The uncertainty principle for energy and time implies that this transition cannot occur 
at an exact time, except in the limit of infinite mean energy~\cite{Hilge}.}
and since exact positions do not exist, neither do exact times. What is true of the positions of 
objects is therefore equally true of the times of events: they have values only to the extent 
that values are indicated.

As the world is differentiated spacewise by the spatial relations (or relative positions) that 
exist in it, so it is differentiated timewise by the temporal relations (or relative times) that 
exist in it. While relative positions exist only as relations between material objects, relative 
times exist only as relations between actual events. The argument by which the world's finite 
spatial differentiation has just been deduced from the nonexistence of exact positions, can 
therefore be repeated almost verbatim to deduce the world's finite temporal differentiation 
from the nonexistence of exact times. But if the world is not infinitely differentiated timewise, 
there can be no such thing as an evolving instantaneous state.

\section{\large THE POINTER PROBLEM}
\label{pp}
A fundamental physical theory that is essentially an algorithm for assigning probabilities to 
VIEs on the basis of other VIEs presupposes the occurrence of VIEs, and the challenge is to 
establish the internal consistency of such a theory. For this reason one of the more frequent 
worries of measurement theorists is to explain how possibilities---or worse, 
probabilities~\cite{Treiman}---become facts, how properties emerge~\cite{JoosZeh}, or 
why events occur~\cite{Pearle}. Yet it isn't the task of a probability theory to account for the 
occurrence of the events to which, and on the basis of which, it assigns probabilities. While 
VIEs are causally linked to the future (inasmuch as they create traces or records, which is 
necessary for their being VIEs), they are causally decoupled from the past, and this not just 
for the trivial reason that the result of a successful measurement is generally not 
necessitated by an antecedent cause. Every quantum-mechanical probability assignment {\it 
presupposes\/} the occurrence of a VIE, and therefore QM (qua probability algorithm) cannot 
supply sufficient conditions for the occurrence of a VIE. If QM (qua probability algorithm) is 
fundamental and complete, it follows that VIEs are uncaused.

Every physical theory has to name those of its ingredients that correspond to reality or 
represent what is. The entire theoretical structure of QM---consisting of its formal structure 
and the ontological structures described by it---must contain a particular substructure that 
can consistently be regarded as factual {\it per se\/}, and that is capable, therefore, of 
accommodating the VIEs presupposed by QM. Our task as measurement theorists is not to 
account for the occurrence of VIEs, let alone for the realization of possibilities unaided by 
VIEs, but to identify this substructure.

Here is how the PIQM carries out this task. The departure of an object~$O$ from a precise 
trajectory can be indicated only if there are detectors that can probe the region over which 
$O$'s fuzzy position extends. This calls for detectors whose position probability distributions 
are narrower than~$O$'s. Such detectors do not exist for all objects. Some objects have the 
sharpest positions in existence. For these objects the probability of a position-indicating 
event (PIE) that is inconsistent with a precise trajectory, is necessarily very low. Hence {\it 
among\/} these objects there are objects of which the following is true: every one of their 
indicated positions is consistent with (i)~every prediction that is based on their previous 
indicated positions and (ii)~a classical law of motion. Such objects deserve to be called 
``macroscopic''.%
\footnote{This definition does not require that the probability of finding a macroscopic object 
where classically it could not be, is strictly zero. What it requires is that there be no PIE that 
is inconsistent with predictions based on a classical law of motion and earlier PIEs.}
To enable a macroscopic object to play the role of a pointer, one exception has to be made: 
its position may change unpredictably if and when it serves to indicate a measurement 
outcome.

Since the indicated positions of macroscopic objects are consistent with classical trajectories, 
we can think of the positions of macroscopic objects (``macroscopic positions'', for short) as 
forming a system of causally connected properties that are effectively detached from the 
events by which they are indicated. We can ignore their dependence on PIEs because their 
effectively deterministic correlations permit us to think of them as a self-contained and 
self-existent causal nexus (interspersed with unpredictable changes in outcome-indicating 
positions).

The cogency of these conclusions hinges on the meaning of ``effectively detached''. 
Macroscopic positions are not really detached from the events by which they are indicated. 
No position is possessed unless its possession is indicated. Even the Moon is where it is only 
because of the (myriad of) events that betoken its whereabouts. Macroscopic positions are 
possessed (to the extent that they are) because they are indicated by macroscopic positions 
(to the extent that they are). This mutual dependence, however, does not affect the 
independent existence of the {\it entire\/} system of macroscopic positions (the 
``macroworld''). While ontologically it would be wrong to consider a single macroscopic 
position as factual {\it per se\/}, nothing stands in the way of considering the entire system 
as factual {\it per se\/}.

Moreover, for all {\it quantitative\/} purposes (FAQP) it is perfectly legitimate to ignore the 
mutual dependence of macroscopic positions and to treat them as {\it individually\/} 
self-indicating. The existence of truth-value-lacking propositions of the form ``$Q$~has the 
value~$q$'' is a consequence of the fuzziness of physical variables. Such propositions call for 
a criterion for the existence of a truth value, and this consists in the occurrence of a VIE 
(indicating both the truth value of a proposition and the value of an observable). The 
fuzziness of physical variables thus implies the supervenience of possessed values on VIEs. 
The fuzziness of a macroscopic position, on the other hand, never evinces itself through 
unpredictable PIEs. (Macroscopic objects, recall, are defined that way.) It only exists in 
relation to an imaginary spatial background that is more differentiated than the physical 
world. It therefore is as unreal physically as the intrinsically and infinitely differentiated 
spatial background of classical theories. This is why it is perfectly legitimate FAQP (rather 
than merely FAPP) to ignore the supervenience of macroscopic positions on PIEs.

The structure that represents what exists, and that contains the VIEs presupposed by QM, 
thus turns out to be what every experimental physicist takes it to be: the system of 
macroscopic positions. It is a substructure in two senses: it is part of the entire theoretical 
structure of QM, and it is the self-existent foundation on which all indicated values 
supervene. It contains the VIEs as unpredictable transitions of value-indicating positions. 
(Since such a position belongs to the self-existent macroworld both before and after a 
value-indicating transition, the transition partakes of the factuality of the macroworld.) What 
makes this structure the sole candidate for the predicate ``factual {\it per se\/}'' is the 
physical unreality of its own fuzziness, existing as it does solely in relation to a nonexistent 
``manifold''. Thus even though there is no hermitian ``factuality operator'' (one cannot 
measure factuality), QM (qua probability algorithm) uniquely determines what is factual {\it 
per se\/}.

What light does this shed on the following ``transition''?
\be\label{e1}
\ket{s}\otimes\ket{m_0}=\sum_i c_i\ket{q_i}\otimes\ket{m_0}\longrightarrow
\sum_i c_i\ket{q_i}\otimes\ket{m_i}.
\ee
$m_i$ is the property whose possession, by an apparatus~$\cal A$, indicates that a 
system~$\cal S$ has the property (or an observable~$Q$ has the value)~$q_i$, and $m_0$ 
is the apparatus-property of being in the neutral state. As it stands, (\ref{e1}) is not a 
physical transition but either a conditional probability measure or a substitution of one 
probability measure for another reflecting a change in the time of the possible measurements 
to the possible results of which probabilities are assigned. Assuming that the possession 
of~$s$ by~$\cal S$ and that of~$m_0$ by~$\cal A$ at the initial time are indicated, the 
final probability measure tells us (among other things) that the probability of finding both 
$m_i$ and~$q_k$ at the final time (given that the appropriate measurements are 
successfully made) is $|c_i|^2$ if $k=i$ and 0 otherwise. If the initial possession of~$s$ 
and~$m_0$ is not indicated, the initial ``state'' is itself a probability measure based on 
earlier VIEs, assigning probability~1 to the joint detection of $s$ and~$m_0$ (given that the 
appropriate measurements are successfully made).

Let us now take into account that $m_0$ stands for the neutral position of a macroscopic 
pointer, and that the possession by $\cal A$ of one of the properties $m_i$ indicates the 
result of a measurement. This means that the initial possession of $m_0$ and the final 
possession of one of the $m_i$ are embedded and recorded in the self-existent system of 
effectively sharp positions that make up the macroworld. In this case the transition of $\cal 
A$ from having the property~$m_0$ to having one of the properties~$m_i$ is a physical 
transition, indicating that $\cal S$~has the corresponding property~$q_i$ at the time of the 
transition.

If the so-called ``measurement problem'' is the problem of how a particular element of some 
decomposition of the final algorithm comes to represent (or to appear to represent) what 
exists, the so-called ``pointer problem'' is the problem of why it is an element of this 
decomposition rather than another. As pointed out at the beginning of the present section, 
the first problem is spurious. All that needs to be established is the consistency of the 
spontaneous occurrence of uncaused VIEs with the quantum-mechanical correlations 
between VIEs, and this has just been done, at least in a qualitative or heuristic fashion. To 
render the argument quantitative, it suffices to invoke the results of decoherence 
investigations.

These investigations generally begin by dividing the world into a ``collective system'' 
(system$\,+\,$apparatus) and its ``environment''. What they demonstrate is that, for a 
large class of models if not in complete generality~\cite{Omnes}, the reduced density 
operator of the collective system, obtained by a partial trace on the environment, becomes 
virtually diagonal with respect to a privileged basis in a very short time, and that it stays that 
way for a very long time. Because all known interaction Hamiltonians contain 
$\delta$-functions of the distances between particles, the privileged basis is that defined by 
the collective position variables.

This quantitatively underpins (i)~the existence of objects for which the probability of a PIE 
that is inconsistent with a precise trajectory is very low, (ii)~the conclusion that there are 
macroscopic objects in the specific sense spelled out above, and (iii)~the consistency of the 
correlations encapsulated by~QM with the assignment of independent existence to the 
system of macroscopic positions. The PIQM thus solves the pointer problem by availing itself 
of the tools of decoherence theories. At the same time it avoids the self-defeating philosophy 
espoused by decoherence theorists; for these take an encapsulation of correlations between 
VIEs, transmogrify it into an evolving state of affairs, and as a consequence are forced to 
deny the objective reality of VIEs: ``While decoherence transforms the formal `plus' of a 
superposition into an effective `and' (an apparent ensemble of new wave functions), this 
`and' becomes an `or' only with respect to a subjective observer''~\cite{Zeh}. (The reason 
why the ``and'' is only ``effective'' is that the reduced density operator never becomes 
completely diagonal: decoherent histories exist only FAPP.)

What decoherence investigations demonstrate is that decoherence makes the environment a 
more accurate monitor of macroscopic positions than any individual apparatus could be. 
Individual measurements of macroscopic positions reveal pre-existent properties, in the 
sense that they indicate properties that are already indicated by the environment. Note that 
while this general result doesn't make much sense in the context of the aforesaid philosophy, 
it makes good sense if one takes account of the supervenience of possessed values on VIEs, 
which follows from the fuzziness of physical variables.

Modal interpretations ``solve'' the problem of how the ``and'' becomes an ``or'' by turning 
it into a postulate: whenever a two-component system has a unique biorthogonal 
decomposition---this is the case in the event that all $c_i$ have different norms---exactly 
one term of this decomposition represents the actual state of the system. There remains the 
problem of how an effective ``or'' becomes a bona fide ``or''. To arrive at a bona fide ``or'' 
within the modal scheme one has to establish the exact biorthogonality of the decomposition 
of the final state in (\ref{e1}), now interpreted as a state of the collective system and the 
environment. The fact that the reduced density operator of the collective system,
\be
\sum_{ij}c_ic_j^*\braket{m_j}{m_i}\ketbra{q_i}{q_j},
\ee
is diagonal (regardless of the values of the~$c_i$) only if the environment kets $\ket{m_i}$ 
are orthogonal, suggests that modal interpretations and decoherence theories are equally 
incapable of establishing a genuine ``or''. In addition it can be held against modal 
interpretations that (i)~they consider an ever so small quantitative difference sufficient for a 
considerable conceptual difference, namely the difference between the existence and the 
nonexistence of a value~\cite{pseudo}, and (ii)~they feature a superfluous postulate. As has 
been shown, QM (qua probability algorithm) is perfectly capable of a unique and consistent 
reality assignment, without extraneous postulates.

In summary, I believe that, contrary to Marchildon's assessment, the PIQM does bring 
additional insight to bear on the pointer problem---additional to what is achieved by 
decoherence theories and modal interpretations.

\section{\large RISK ASSESSMENT}
\label{ra}
According to the PIQM, VIEs are uncaused. Marchildon considers this conclusion 
epistemologically ``risky''. He believes that classical mechanics explains everything except 
the initial conditions: ``the totality of the world at one instant is unexplained, but the totality 
of the world at all other instants is explained''. He goes on to say that, according to the 
PIQM,
\bq
all the facts that constantly betoken positions of macroscopic objects should be taken as 
unexplainable. It seems that the unexplained here far exceeds what it is in classical 
mechanics. Inevitably, people will look for regularities in the occurrence of facts\dots. In a 
sense, spontaneous localization theory may be viewed as doing just that.\ \dots any theory 
that would correctly account for the occurrence of facts would, other things being equal, have 
a head start over one that does not.
\eq
Let us examine these claims. To begin with, the quantum-mechanical correlations between 
macroscopic objects are deterministic, in the sense that every indicated macroscopic position 
is consistent with every prediction that is based on previous indicated macroscopic positions 
and a classical law of motion. This makes it possible to think of them as correlations between 
causes and effects, rather than as correlations between VIEs or values indicated by VIEs, and 
to forget about the supervenience of macroscopic positions on VIEs, at least FAQP. It would 
be delusional, however, to believe that the deterministic correlations of classical mechanics 
are thereby explained.

Take classical electrodynamics. It is an algorithm for calculating, in two steps, the effects 
that electrically charged objects have on electrically charged objects. Given the distribution 
and motion of charges, we calculate four or six functions of spacetime coordinates, the 
components of either the 4-vector potential or the electromagnetic field. Given these 
functions, we calculate the effect that those charges have on other charges (or would have if 
other charges were present). If there is an underlying process or mechanism by which 
charges act on charges, we know nothing of it. The transmogrification of an algorithm for 
calculating effects into a process or mechanism by which these effects are produced is but a 
sleight of hand. Besides, the notion that the electromagnetic field is a physical entity in its 
own right, which is locally acted on by charges, locally acts on charges, and locally acts on 
itself, still doesn't explain how the field is locally acted on by charges, how it locally acts on 
charges, and how it locally acts on itself.%
\footnote{One is left to wonder why we tend to stop worrying once we have transformed the 
mystery of action at a distance into the mystery of local action. Perhaps it is because 
``physicists are, at bottom, a naive breed, forever trying to come to terms with the `world 
out there' by methods which, however imaginative and refined, involve in essence the same 
element of contact as a well-placed kick''~\cite{DeWGra}.}

Again, classical mechanics cannot account for the stability of matter---the existence of 
objects that are composed of finite numbers of particles, that ``occupy'' finite regions of 
space, and that do neither collapse nor explode as soon as they are created. QM does. The 
stability of matter hinges on the fuzziness of the internal relative positions and momenta of 
composite objects, which finds its quantitative expression in the statistical correlations of QM. 
Does this justify the claim that (according to the PIQM) the unexplained ``far exceeds what 
it is in classical mechanics''? I don't think so.

Next, to say that ``all the facts that constantly betoken positions of macroscopic objects 
should be taken as unexplainable'' is to overstate the lack of sufficient conditions for the 
occurrence of a VIE. Many aspects of the click of a counter or the deflection of a pointer can 
be understood in terms of the deterministic correlations that structure the macroworld. As 
Marchildon correctly points out, ``properties of facts may be subject to experimental 
investigation and, therefore, call for theoretical analysis''. Theoretical analysis is an analysis 
of types and causal relations. The absence of causally sufficient conditions for the occurrence 
of an event of type~X does not preclude a detailed theoretical analysis of events of type~X. 
To say that facts are irreducible, as Marchildon does, is therefore incorrect. What is 
irreducible is only the factuality of facts and the {\it occurrence\/} of value-indicating facts.

As Marchildon's allusion to spontaneous localization theory suggests, if the lack of causally 
sufficient conditions is a drawback, it is a drawback of standard QM (unadulterated by 
spontaneous localizations or such) rather than a drawback of the PIQM.%
\footnote{It seems to me that it would take a far more substantial overhaul of QM than a 
nonlinear or stochastic modification of the dependence of probability assignments on the 
times of VIEs, in order to arrive at a theory featuring causally sufficient conditions for VIEs. 
The efficiency of any actually existing detector is related to the value of at least one coupling 
constant. This contributes to determine the probability of detection in the event that the 
(Born or ABL) probability of detection is~1. By building large redundancies into our detectors 
(which is possible only for large detectors) we can make this probability close to~1, but we 
cannot make it exactly~1. And if it is not exactly~1, the detection event, if it occurs, cannot 
be said to have been necessitated (caused).}
But is it a drawback? One thing is certain: The inability to formulate sufficient conditions for 
VIEs does not imply that QM is incomplete, for it could be the physical world that is 
``incomplete''. If QM is fundamentally a probability algorithm quantifying (among other 
things) the fuzzy spatial relations that contribute to ``fluff out'' matter, this is indeed the 
case: the physical world does not contain all the spatiotemporal distinctions that we tend to 
make, and its spatiotemporal differentiation does not go all the way down. If QM's being 
fundamentally a probability algorithm entails the impossibility of formulating sufficient 
conditions for VIEs, this is a small price to pay for a significant insight into the spatiotemporal 
structure of the physical world. Moreover, the nonexistence of causally sufficient conditions 
for VIEs is itself (a direct consequence of) another significant insight: The applicability of 
causal concepts is confined to the macroworld. Causal ``explanations'' are causal 
interpretations of statistical correlations, and such interpretations are possible only in the 
limiting case in which the correlations become deterministic in the sense just spelled out 
(that is, only within the system of macroscopic positions, which evolves deterministically 
except for the value-indicating transitions that occur in it).

It may be true that ``any theory that would correctly account for the occurrence of facts 
would, other things being equal, have a head start over one that does not'', only there is no 
such theory. There are always relevant ``other things'' that are not equal. As d'Espagnat has 
argued at length and convincingly~\cite{dE1,dE2,dE3}, one has either to accept a modified 
``dynamics'' or content oneself with a theory that is objective only in a weak, intersubjective 
sense, inasmuch as it countenances a conceptual fuzziness that allows ``$\approx$'' to do 
duty for~``$=$''~\cite{pseudo}.%
\footnote{Even spontaneous localization theories are not free from this kind of fuzziness, 
since the collapsed probability distributions have ``tails''.}
The PIQM has to do neither because it does not share the assumption on which d'Espagnat's 
conclusion rests. By rejecting the assumption that a quantum state is an evolving 
(instantaneous) state of affairs of some sort, as well as the intrinsically and infinitely 
differentiated background time/spacetime that this assumption entails in a 
nonrelativistic/relativistic context, the PIQM avoids the conceptual fuzziness of statements 
that are true only FAPP. It features, instead, an objective fuzziness and statements that are 
true FAQP.

\section{\large THE PSYCHOLOGY OF QUANTUM STATE\\
EVOLUTION}
Marchildon adduces a couple of ``good'' (albeit not compelling) reasons to view the state 
vector as specifying an evolving (instantaneous) state. The first is this: Almost everybody 
agrees that a complete measurement performed on a system~$\cal S$ at a time~$t_1$ 
warrants the inference that at this time $\cal S$ possesses a property (or a set of properties) 
that can be represented by a one-dimensional projector~$\bfP(t_1)$. Knowing this property 
as well as the system's Hamiltonian, we can predict that the appropriate measurement, if 
successfully performed at a time $t_2>t_1$, will warrant (or would warrant) the inference 
that $\cal S$ possesses the property represented by $\bfP(t_2)$ at the time~$t_2$. This is 
usually considered sufficient ground for the belief that $\cal S$ possesses $\bfP(t_2)$ 
at~$t_2$ even if it is not the outcome of an actually performed measurement. ``This 
possibility of correctly predicting the value of a time-dependent dynamical variable motivates 
the association of the state vector with an actual state of affairs''~\cite{Marchildon}.

It is, however, equally possible to correctly {\it retrodict\/} the value of a time-dependent 
variable. Knowing $\bfP(t_1)$ and the system's Hamiltonian, we can retrodict that the 
appropriate measurement, if successfully performed at a time $t_2<t_1$, did warrant (or 
would have warranted) the inference that $\cal S$ possesses $\bfP(t_2)$ at~$t_2$. But this 
means that Marchildon's first ``good'' reason is a good reason {\it not\/} to view the state 
vector as specifying an (instantaneous) state that evolves (toward the future). All 
quantum-mechanical probability assignments are time symmetric. Born probabilities can be 
assigned on the basis of either past or future VIEs, and ABL probabilities~\cite{ABL} are by 
nature time-symmetric, whereas the notion of an evolving state of affairs is anything but 
time-symmetric.

The second ``good'' reason is the belief, largely due to von Neumann~\cite{vN}, that a 
measurement is an interaction followed by a collapse. Marchildon considers a series of three 
measurements---of $A$ yielding $\ket{a}$ at~$t_1$, of $B$ yielding $\ket{b}$ at a later 
time~$t_2$, and of $C$ yielding $\ket{c_i}$ at an intermediate time~$t$---and argues that
\bq
For the purpose of computing probabilities of measurement results at time~$t$, 
state~$\ket{a}$ and state~$\ket{b}$ are equally useful. For the purpose of making 
ontological statements, however, it is more natural in the absence of a $C$~measurement to 
hold that~$\ket{a}$, rather than~$\ket{b}$, is the intermediate state, since $\ket{b}$ 
obtains only after an intervening interaction.
\eq
According to von Neumann, (i)~$\ket{b}$~only obtains after an intervening interaction, and 
(ii)~the state vector specifies an evolving (instantaneous) state of affairs. If one makes 
assumption~(ii) then it is natural to attribute to the state vector two modes of evolution, a 
continuous one that (owing to some interaction between system and apparatus) leads to an 
entangled state according to formula~(3), and a subsequent ``collapse''. Evidently, 
$\ket{b}$ then obtains only after the collapse. Conversely, if one makes 
assumption~(i)---that is, if one regards $\ket{b}$ as due to a collapse from an entangled 
state brought about by a prior interaction---then it is natural to look upon the state vector as 
an evolving (instantaneous) state of affairs. The two assumptions support each other. What 
is missing is a good reason that supports them both.

It is indeed ``more natural'' to hold that $\ket{a}$ is the intermediate state, but for reasons 
that are psychological rather than physical. We are accustomed to the idea that the redness 
of a ripe tomato exists in our minds, rather than in the physical world. We find it 
incomparably more difficult to accept that the same is true of the experiential now: it has no 
counterpart in the physical world. There simply is no objective way to characterize the 
present. And since the past and the future are defined relative to the present, they too 
cannot be defined in physical terms. The temporal modes past, present, and future can be 
characterized only by how they relate to us as conscious subjects: through memory, through 
the present-tense immediacy of qualia, or through anticipation. In the physical world, we 
may qualify events or states of affairs as past, present, or future {\it relative to\/} other 
events or states of affairs, but we cannot speak of {\it the\/} past, {\it the\/} present, or 
{\it the\/} future. The idea that some things exist not yet and others exist no longer is as 
true (psychologically speaking) and as false (physically speaking) as the idea that a ripe 
tomato is red.

If we conceive of temporal or spatiotemporal relations, we conceive of the corresponding 
relata simultaneously---they exist at the same time {\it in our minds\/}---even though they 
happen or obtain at different times in the physical world. Since we cannot help it, that has to 
be OK. But it is definitely not OK to think of the spatiotemporal whole as a simultaneous 
spatial whole that persists in time, and to imagine the present as advancing through it. There 
is only one time, the fourth dimension of the spatiotemporal whole. There is not another time 
in which this spatiotemporal whole persists as a spatial whole and in which the present 
advances. If the experiential now is anywhere in the spatiotemporal whole, it is trivially and 
vacuously everywhere---or, rather, everywhen.

In a world that has no room for an advancing now, time does not ``flow'' or ``pass''. To 
philosophers, the perplexities and absurdities entailed by the notion of an advancing 
objective present or a flowing objective time are well known. (See, e.g., the illuminating 
entry on ``time'' in Ref.~\cite{Audi}.) To physicists the nonexistence of an advancing 
present was brought home by the discovery of the relativity of simultaneity. This was 
Nature's refutation of ``presentism'', the view that only the present is real. The opposite 
view, that no time is ``more real'' than any other time, however, is so counterintuitive 
(given the distinctiveness of the experiential now), that few physicists can resist the fallacy 
of projecting an advancing now into the physical world.%
\footnote{Those who can, tend to go to the other extreme of altogether denying the 
temporality of the physical world, e.g.: ``The objective world simply {\it is\/}, it does not 
{\it happen\/}\dots''~\cite{Weyl}.}

In the maximally differentiated world of classical physics this leads to the well-known folk 
tale according to which causal influences reach from the nonexistent past to the nonexistent 
future through persisting ``imprints'' on the present. If the past is unreal, it can influence 
the (equally unreal) future only through the mediation of something that persists. Causal 
influences reach from the past into the future by being ``carried through time'' by something 
that ``stays in the present''. There is, accordingly, an evolving instantaneous state, and this 
includes not only all presently possessed properties but also traces of everything in the past 
that is causally relevant to the future. This is how we come to conceive of fields of force that 
evolve in time (and therefore, in a relativistic world, according to the principle of local 
action), and that mediate between the past and the future (and therefore, in a relativistic 
world, between local causes and their distant effects). Classical electrodynamics is a case in 
point. What compels one to transmogrify this algorithm for calculating the effects that 
electrically charged objects have on electrically charged objects (Sec.~\ref{ra}) into a local 
mechanism or a continuous process by means of which effects are produced, is the attempt 
to foist an advancing now into the world of classical physics. The attempt to foist an 
advancing now into the quantum world compels one to seize instead on a probability 
algorithm and to transmogrify the same into an instantaneous state that plays a similar 
mediating role.

The are other reasons for not viewing the state vector as specifying an evolving 
(instantaneous) state. One is the relativity of simultaneity. Simultaneity is a feature of the 
language by which we describe the world, rather than a feature of the world. Delocalized 
evolving quantum states have the nasty habit of collapsing simultaneously, which means that 
collapses, too, are features of our language rather than features of the world. ``Collapse is 
something that happens in our description of the system, not to the system 
itself''~\cite{FuPer}. A description that introduces features that correspond to nothing in the 
physical world may be convenient for certain purposes, but for making ontological 
statements a coordinate-free description is obviously superior. Since simultaneity does not 
feature in such a description, neither can collapses. It follows either that quantum states are 
not evolving instantaneous states, or that they are such states but never collapse, in which 
case we have a theory that reifies a mathematical encapsulation of correlations between 
events the reality of which it denies (Sec.~\ref{pp}).%
\footnote{And what about GHZ correlations~\cite{GHSZ}? The values of all three spin 
components of each of the three particles can be correctly predicted on the basis of 
measurements performed on the other two particles. Does the possibility of correctly 
predicting these values warrant the association of the state vector with an actual state of 
affairs? It certainly does not warrant the association of values with unmeasured spin 
components, for if one associates values with all $x$ and all $y$~components, the 
correlations imply that the product of the three $x$~components equals {\it both\/} $+1$ 
{\it and\/} $-1$~\cite{Vaidman}.}

Finally, if one takes one's cue from the stability of matter, which hinges on the fuzziness of 
relative positions and momenta, and looks for an appropriate tool for dealing with fuzzy 
variables, one finds this to be a probability algorithm. Since the most direct consequence of 
the fuzziness of physical variables (apart from the very existence of stable objects) is the 
unpredictability of measurement outcomes, the obvious way to quantify a fuzzy variable is to 
assign nontrivial probabilities. What is more, the quantum formalism can be derived by a 
straightforward generalization of the classical probability algorithm~\cite{JustSo}, which 
represents possible outcomes by subsets and probability measures by points 
(Sec.~\ref{hqse}).  To make room for nontrivial probabilities, one represents probability 
measures by 1-dimensional rays instead of 0-dimensional points, and possible outcomes by 
subspaces instead of subsets. The rest follows via Gleason's theorem.

\section{\large ON THE ``TIME OF MEASUREMENT''}
Marchildon asks, ``Is a detector required to have some minimum mass to be able to work as 
a detector?'' Much of the answer depends on what one means by a ``detector''. What I mean 
is an object capable of indicating the presence of another object in a (more or less 
well-defined) region of space that exists by virtue of being defined in terms of self-existent 
macroscopic positions---the positions (say) of the macroscopic parts of the material 
boundary of the detector's sensitive region. (The reason why this region is only more or less 
well defined is that its boundary is eventually made up of non-macroscopic objects with 
measurably fuzzy positions.) In addition, a detector must be capable of indicating the 
presence of another object in its sensitive region, and this requires a macroscopic pointer. 
Without a macroscopic pointer there is nothing that can indicate the possession of an 
attributable position, and without a region defined in terms of macroscopic positions, there is 
no attributable position (Sec.~\ref{fstd}).

Given my definition of ``macroscopic'' (Sec.~\ref{pp}), an object with a large mass is very 
likely to be macroscopic, and an object with a small mass is unlikely to be so. There is 
therefore no precise mass limit at which an object ceases to be macroscopic. By the same 
token, there is no precise mass limit below which an object cannot function as a detector. 
Marchildon raises the question of whether an atom can work as a detector. Perhaps the 
atomic tip of a scanning tunneling microscope can be regarded as a detector (for electrons 
with energies matching the potential between the tip and the sample), but only because it is 
attached to a macroscopic object and therefore in possession of a rather precise position, and 
only because a macroscopic pointer indicates the rate at which electrons are detected. 
Without the benefit of such macroscopic paraphernalia, no atom can work as a detector. 
Marchildon next considers a hydrogen atom in a $2p$~state, confined in a high vacuum.
\bq
In a split second the atom is very likely to emit a 1216\thinspace\AA\ photon and fall to the 
ground state. Perhaps one will maintain that when the photon has gone far enough, a fact 
has occurred. But then the question is, When did the fact occur?\dots If it is agreed that the 
fact has already occurred at $t=1\,$s\dots, it must have occurred at some instant between 
$t=0$ and $t=1\,$s. That instant is defined as the one starting from which it is impossible to 
undo or erase.
\eq
To begin with, since even probability~1 is not sufficient for ``is'' or ``has'' 
(Sec.~\ref{fstd}), the fact that the probability of finding the atom in its ground state is very 
close to~1 does not warrant the conclusion that the atom is in its ground state. Unless either 
the atom is found in its ground state or (what usually comes to the same) the photon is 
detected, no fact has occurred.

Next suppose that both the atom's being in a $2p$~state at $t=0$ and its being in the 
ground state at $t=1$ are indicated. From this it does not follow that the transition has 
occurred at some instant between $t=0$ and $t=1$. If all that can be inferred from the 
goings-on in the macroworld is the atom's respective states at these times then saying that 
the transition has occurred between $t=0$ and $t=1$ means no more than that the atom 
was in a $2p$~state at $t=0$ and in the ground state at $t=1$. In this case the state of 
affairs that obtains in the meantime is a temporally undifferentiated state of affairs 
(Sec.~\ref{hqse}).

Marchildon defines the transition time as the earliest time after which it is impossible to undo 
or erase the fact by which the transition is indicated. In point of fact, there is no such time, 
for it is strictly impossible to undo or erase a VIE. In order to be a VIE, it must create a 
record, with the help of a value-indicating position that is embedded in the self-existent 
macroworld. Because the latter evolves deterministically, it retains information about the 
indicated value. Thus before there is a record, there is no VIE, and once there is a VIE, there 
is a record that can be erased only FAPP.%
\footnote{The so-called ``quantum erasure'' of information~\cite{gy,esw,sew} is not an 
erasure of macroscopically recorded information but merely an ``erasure'' of the possibility 
of obtaining such information.}

The notion that a VIE can be erased stands or falls with the notion that a quantum state has 
a kind of reality of its own, distinct from the reality of actual events and states of affairs 
(including measurements). One may see the quantum state as an evolving matrix of 
``propensities''~\cite{Popper} or ``potentialities''~\cite{Heisenberg, Shimony}, or one may 
look upon it as the {\it real\/} reality, to be distinguished from the apparent reality 
experienced by information gathering and retaining system like us~\cite{JoosZeh,Zurek}. In 
either case one searches the reality of one kind for criteria that warrant the other kind of 
reality. If one then believes that a particular kind of entanglement in the reality of the first 
kind (e.g., entanglement with a sufficiently massive object~\cite{gy}, or entanglement that 
is FAPP irreversible) warrants a reality assignment of the second kind, the possibility of 
reversing such a reality assignment exists. But only then.

Marchildon's remarks nonetheless touch on an ambiguity concerning the time of a VIE that 
ought to be cleared up. A VIE lacks a causal precursor, and it leaves a trace that can be 
erased only FAPP. The paradigm example of a VIE---the deflection of a loudspeaker 
membrane or a pointer needle---satisfies the requirement of being embedded in the 
macroworld, but it does not seem to lack a causal precursor. After all, a detector does not 
click without (say) an initial ionization, and the paradigmatic click is no more than a 
convenient way to refer to the causal chain by which information is retained in the 
deterministically evolving macroworld. The converse, however, is equally true: without the 
click or record there would be no initial ionization. Like the measurement outcome indicated 
by the click, the initial ionization event supervenes on the goings-on in the macroworld.

This necessitates a distinction between the causality we mentally project into the 
macroworld, onto the deterministic correlations between macroscopic positions, and the 
causality by which we extend a record backward as far as counterfactuals permit. The first is 
an (imaginary) connection between events of which there are separate records. The second 
connects an individual record to the property or value that is indicated by the record and 
supervenes on it. There is no harm in regarding an initial ionization as causing the click (and 
an incoming particle as causing the ionization) on the strength of a counterfactual---without 
an incoming particle there would have been no ionization, and without ionization there would 
have been no click---as long as we remember the flip side: without the click or record there 
would have been no ionization, and without an ionization event, there would have been no 
incoming particle.

As we extend our story towards the past, it becomes less and less distinct. The question of 
which atom in the counter was the first to get ionized, for instance, involves distinctions that 
Nature does not make. Still less can be inferred regarding the incoming particle. If the setup 
warrants it, we can identify this particle with a particle detected earlier elsewhere, but even 
in this rather exceptional case the causal chain cannot be extended beyond the first 
ionization by the incoming particle. The particle's indicated presence here and now is 
certainly not a consequence of the particle's indicated presence there and then.

The time of a position measurement, accordingly, is not the time of a click (which, after all, 
just stands for an enduring record) but the time of the first event in a causal chain. As far as 
time is concerned, this chain begins (in our example) with the first ionization of an atom by 
an incoming particle. From an ontological point of view, however, the chain begins with an 
unpredictable change in the value of a macroscopic position. The earlier part of the chain 
supervenes on the later part of the chain.%
\footnote{This supervenience has significant implications for cosmology. As we approach the 
cosmological time $t=0$ (from later times), we enter an era in which there is as yet no 
macroworld. About this era we can make two kinds of statements: statements that are true 
(or false) only because their truth values are indicated much later, and counterfactuals. If we 
assign a density operator to this era, it is an ``advanced'' or ``retropared'' density operator 
rather then a ``retarded'' or ``prepared'' one. It ``evolves'' from later VIEs toward the past 
in the same (spurious) sense in which a prepared density operator evolves from earlier VIEs 
futurewards~\cite{iucaa}.}

\section{\large HOW QUANTUM SYSTEMS EVOLVE}
\label{hqse}
I concede to Marchildon that it is not logically inconsistent, as I have 
claimed~\cite{18errors}, to interpret an algorithm for assigning probabilities to the possible 
outcomes of possible measurements (on the basis of actual measurement outcomes) as an 
evolving state of affairs. Classical mechanics does indeed provide a counterexample. Its 
probability measure, a point~$\cal P$ in a phase space, assigns trivial probabilities to all 
possible measurement outcomes, which are represented by subsets---probability~1 to 
subsets containing~$\cal P$, and probability~0 to subsets not containing~$\cal P$. Because 
the probability measure is trivial, it can be interpreted as an evolving state of affairs. If the 
only probabilities are 1~and~0, nothing stands in the way of interpreting probability~1 as 
``has'' and probability~0 as ``lacks''~\cite{JustSo}.

Let us then ask: What changes in our concepts of ``state'' and ``evolution'' are needed so 
as to make them applicable to a world that is fundamentally and irreducibly described by the 
probability distributions of~QM? To begin with, there is no instantaneous state. There are 
states that obtain at the times of measurements, and then there are states that obtain 
between measurements. There are two reasons why the times of measurements are fuzzy. 
For one, they are indicated by fuzzy positions, notwithstanding that macroscopic positions 
are fuzzy only in relation to an imaginary spatiotemporal background that is more 
differentiated than the physical world. For another, the time of possession of an indicated 
property (or the time of the onset of a property-indicating causal chain) supervenes on 
macroscopic events in this chain (e.g., the click of a counter). In general the time of the click 
does not sharply determine the time of possession; this adds to the fuzziness.

At the time $t_a$ of a complete measurement, system~$\cal S$ possesses a property that is 
represented by a one-dimensional projector~${\bf P}_a$. Let us assume that $\cal S$ is not 
subjected to any further measurement. Then all we have to describe the state of affairs that 
obtains thereafter is~${\bf P}_a$. This describes a fuzzy state of affairs in terms of 
probability distributions over the possible results of unperformed measurement.

If the Hamiltonian is not~0 then the probability distributions describing the subsequent fuzzy 
state of affairs depend on the times of unperformed measurements. This is not the same as 
saying that the subsequent fuzzy state of affairs changes with time, for the antecedents of 
these counterfactual probability assignments are false not only because they affirm that a 
measurement is made but also because they affirm that this is made at a particular time. 
Where $\cal S$ is concerned, there is no particular time until another measurement is made. 
$\cal S$~is temporally differentiated by its VIEs. Between actual measurements it is only 
counterfactually differentiated (by unperformed measurements).

If another measurement is subsequently made, the fuzzy state of affairs that obtains in the 
meantime (during which no measurement is made) is fully described only if all relevant 
information is taken into account. This includes the result of the subsequent measurement. 
(Probability assignments based on earlier and later measurement data are made according to 
the ABL rule~\cite{piqm,ABL}.) Thus if, instead of filling the unmeasured gaps between 
measurements with stories that are extraneous to QM (qua probability algorithm), we let QM 
(qua probability algorithm) say all that it is capable of saying, what obtains between 
measurements is what is appropriately described by counterfactual probability assignments, 
namely, fuzzy states of affairs. Like the probability algorithm that describes it, a fuzzy state 
of affairs is not something that evolves. It is not only fuzzy but also temporally 
undifferentiated. What may be said to evolve is quantum {\it systems\/}. The evolution of 
$\cal S$ consists in an alternating succession of indicated states that obtain at the times of 
measurements and temporally undifferentiated fuzzy states that obtain between 
measurements. This is in sharp contrast with the manner in which macroscopic positions 
evolve, whose indicated values merge into continuous histories that are fuzzy only relative to 
a nonexistent spatiotemporal background. (For further discussion see \cite{pseudo}.)

\section{\large CONCLUSION}
Perhaps the most unusual feature of the PIQM, given our tendency of explaining things by 
taking them to pieces, is the supervenience of the microscopic on the macroscopic. To be 
able to understand it, one must conceive of space as the totality of the more or less fuzzy 
spatial relations (relative positions and relative orientations) that hold between material 
objects, rather than as a self-existent and intrinsically differentiated expanse. Fuzzy positions 
have values only to the extent that values are indicated. Values (in this case, spatial regions) 
can be indicated only if they exist---as the sensitive regions of detectors. This, as well as the 
fact that every probability algorithm presupposes the events to which it assigns probabilities, 
suggests that Niels Bohr was right to insist that quantum physics presupposes classical 
physics. But how can the PIQM take QM ``to be fundamental and complete'' and at the same 
time require ``the validity of classical mechanics for its formulation'', as Marchildon affirms?

It is a question of finding the right reality assignment. The fuzziness of a position can evince 
itself only to the extent that less fuzzy positions exist. Since the fuzziness of the least fuzzy 
positions has no observable consequences, it can FAQP be ignored. This makes the system of 
macroscopic positions the only structure to which independent reality can be attributed 
consistently, and it permits us FAQP to treat each macroscopic position as factual {\it per 
se\/} (Sec.~\ref{pp}). If one wants to speak of an ``emergence'' of the ``classical domain'' 
from the ``quantum domain'', this is it. It is, however, a purely theoretical emergence, for 
what emerges is what exists by itself. Instead of being a substrate from which the classical 
domain emerges, the quantum domain supervenes on the classical domain. This is how the 
PIQM fleshes out the mutual dependence of the quantum and classical domains that is 
sometimes invoked (e.g.,~\cite{LL}). It is also how the claim that QM is fundamental and 
complete can be reconciled with the dependence of the quantum domain on the macroworld.

\vspace{\baselineskip}{\bf\noindent Acknowledgment}

\medskip\noindent I wish to thank Louis Marchildon for a stimulating exchange of views.

\end{document}